# Effect of Zr and Al Addition on Nanocluster Formation in Oxide Dispersion Strenghthened Steel – an *ab initio* Study.

Sruthi Mohan*, Gurpreet Kaur, Binaya Kumar Panigrahi, Christopher David, Gangavarapu Amarendra

Material Science Group, Indira Gandhi Centre for Atomic Research, HBNI, Kalpakkam, Tamil Nadu, India - 603102.

**Abstract**

Conventional Oxide dispersion strengthened steels are characterized by thermally stable, high density of Y–Ti–O nanoclusters, which are responsible for their high creep strength. Ti plays a major role in obtaining a high density of ultrafine particles of optimum size range of 2-10 nm. In Al-containing ODS steels developed for corrosion resistance, Y–Al–O clusters formed are of size range 20 -100 nm, and Ti fails in making dispersions finer in the presence of Al. Usage of similar alloying elements like Zr in place of Ti is widely considered. In this study, binding energies of different stages of Y–Zr–O–Vacancy and Y–Al–O–Vacancy complexes in the bcc Iron matrix are studied by first-principle calculations. It is shown that in all the stages of formation, Y-Zr-O-Vacancy clusters have higher binding energy than Y-Al-O-Vacancy clusters and hence in ferritic steel containing both Zr and Al, Y–Zr–O–Vacancy clusters are more stable and more favored to nucleate than Y–Al–O–Vacancy clusters. The bonding nature in each stage is analyzed using charge density difference plots for the plausible reason for higher stability of Y–Zr–O–Vacancy clusters.

*sruthi@igcar.gov.in

## I. Introduction:

Oxide dispersion strengthened steels with radiation-resistant, thermally stable nano-sized dispersions are well known for its resistance to both irradiation and high temperature creep and therefore are regarded as the most promising structural materials for nuclear applications. The dispersions in ODS steels - Y-Ti-O nanoclusters- play an important role in alleviating the accumulation of point defects and trapping of helium. A typical composition of ODS steel is Fe with 0.2 – 0.5 wt% of $Y_2O_3$, 0.1 -0.4 Ti, 9-16 wt% of Cr and 1-3 wt% of W[1]. Y and Ti, along with oxygen forms high-density nanoclusters, alloying with Cr gives corrosion resistance and low activation W provides solid solution strengthening. The performance of ODS ferritic steel in a corrosive environment can be improved by the addition of 4 – 6 wt% of Al to it owing to the formation of a continuous protective layer of $Al_2O_3$[2-6]. In Al free ODS alloys, the protective oxide layer is a monolayer composed of $Cr_2O_3$ or $(Cr, Fe)_2O_3$. With Al-containing alloys, there is a more complex multilayer protective film, with an outer Fe rich oxide layer and inner Al-rich $Al_2O_3$ layer[7]. The addition of Al also changes the chemistry of dispersoids. It is found that initially titanium-rich Y- Ti – O complexes, which are primarily $Y_2Ti_2O_7$ or $YTiO_5$ gets replaced by Y – Al – O complexes on Al addition. The Y- Al – O phases were coarse with a size range 20 -100 nm compared to fine Y- Ti- O phases with a size range 2 –10 nm. Because the dispersoids are larger, the ultimate tensile strength of Al contacting alloy is lower[4]. In Fe – 16 Cr – 4Al – 2W-0.3Ti-0.3 $Y_2O_3$ (Extrusion temperature 1150°C) and Fe- 20 Cr-4.5Al-0.34Ti-0.5$Y_2O_3$ (Extrusion Temperature 1107°C) model ODS alloys, the Al-containing phases are identified as Yttrium Aluminium Monoclinic, YAM ($Y_4Al_2O_9$) and $YAlO_3$. And there is a proposed three stage formation mechanism for Y-Al- O phases, which include: (i) Dissolution of $Y_2O_3$, $Y_2O_3 \rightarrow 2[Y]+3[O]$ (ii) Internal oxidation reaction of Al due to high oxygen affinity of Al (Al > Ti > Cr >Fe), $2[Al]+3[O] \rightarrow Al_2O_3$ and (iii) an oxide formation reaction which can take place at temperatures between 900°C and 1100°C. Studies on Al-containing ODS steels of composition Fe–15.5Cr–2W–0.1Ti–4Al–0.35$Y_2O_3$ by high-resolution transmission electron microscopy and diffraction contrast techniques including weak beam electron microscopy concluded the presence of yttrium–aluminium–hexagonal (YAH, $YAlO_3$) and yttrium–aluminium–perovskite (YAP, $YAlO_3$) oxides. In all the studies, including the ones mentioned above, the number density of $Y_2Ti_2O_7$ or $Y_2TiO_5$ was negligible compared to Y- Al-O phases. And it points to the fact that Ti plays an insignificant role in forming oxide nanoparticles in the presence of Al

[8]. Further, it was found that the addition of a small amount of Zr or Hf results in a significant increase in creep strength and yield strength at elevated temperatures in Al-added ODS steels, which may be due to decrease in precipitate size[2, 9, 10]. It is also observed in TEM and HRTEM studies of Al-alloyed high Cr-oxide dispersion strengthened steel with Zr addition that almost all the small nanoparticles (diameter <10 nm) were found to be consistent with trigonal δ- $Y_4Zr_3O_{12}$ and coherent with the bcc steel matrix, with semi-coherent orthorhombic $Y_2TiO_5$ oxides occasionally detected. The large particles were mainly identified as tetragonal or cubic $ZrO_2$ oxide. Al-containing phases were completely absent[11]. These studies sum up the possibility of Zr as an alternative to Ti in ODS steels containing Al.

Several detailed first-principle calculations are being carried out for in-depth understanding of the structure and stability of nanoclusters in the Iron matrix. One insightful yet disputed study was by C. L. Fu *et al* which pointed out that Oxygen has high formation energy (and low solubility) in defect-free Fe lattice, but the formation energy of O-vacancy pairs becomes surprisingly small if the vacancies pre-exist in the matrix. So he concluded that vacancies are the most crucial alloying element for stable nanoclusters[6]. The role of minor alloying elements like Ti and Zr on the stability and dispersion of nanoclusters has been studied in detail[12]. The binding energies of Y-O-Vacancy clusters increase when Ti is replaced with Zr leading to higher stability. The higher stability of clusters enhances the nucleation rate, which produces finer dispersions[13]. The argument by C. L. Fu was contradicted by A. Claisse *et al.* telling that Y- Ti-O clusters do not need vacancy to stabilize in the bcc Fe matrix and the repulsion between Y and Ti is overcome by their strong attraction to oxygen[14]. Still reported values of Vacancy – Oxygen binding is more than solute – Oxygen binding. Mechanical alloying, by which most of the ODS steels are synthesized also introduces a supersaturation of vacancies. Positron annihilation spectroscopy studies have found vacancy clusters of vacancies inside Y – Ti – O complexes in bcc Fe[15, 16]. These studies together propose a possibility of nucleation of nanoclusters starting from the O-enriched solute clustering with or without vacancy mediation.

This has been experimentally verified by HRTEM studies on Fe -15Cr–2W–0.1Ti–4Al–0.63Zr–0.35$Y_2O_3$ alloy [11]. However, the reason for that has to be studied in depth. Here in this work, we are comparing the energetics of Y-Al –O and Y –Zr-O cluster formation in Fe matrix by DFT calculations to find out which one among these is more stable. The behavior of Zr and Al with Fe matrix containing Oxygen, Vacancy, Oxygen–Vacancy, and Yttrium –Oxygen –Vacancy are evaluated using binding energy values and charge density difference plot to understand basic interactions involved in nanocluster formation in Fe matrix.

## II. Details of Calculation

The calculations are done using the Vienna ab initio simulation package (VASP) within the density functional theory[17, 18]. All the calculations are done with pseudopotentials generated with the projected augmented wave approach (PAW)[19]. For exchange-correlation functional, the generalized gradient approximation as parameterized by Perdew, Burke and Ermzerhof (PBE)[20] is used. A plane wave energy cut off of 500 eV is found to give total energy convergence of better than 0.01 eV/atom. Brillouin-zone integration is done using a 2×2×2 Monkhorst-Pack[21] mesh for the supercell of bcc Fe containing 128 atoms. All the calculations are spin-polarized. The simulation cell is fully relaxed (atom positions and cell volume) until the forces on each atom is less than 0.001 eV/Å.

The formation energy or mixing energy of a substitutional impurity X in bcc iron supercell *of* size N atoms is[22]:

$$E_X(Fe) = E\big((N-1)Fe + 1X\big) - (N-1)E(Fe) - E_{ref}(X), \quad X = Y, Al, Zr \quad \text{---(A)}$$

Where, $E_{ref}(X)$ is the reference energy calculated with respect to equilibrium phases. (fcc for Al, hcp for Y and Zr). If formation energy is positive, the process is endothermic and if it is negative, the process is exothermic. Elements with high positive formation energy are expected to have low solubility in the Fe matrix.

Similarly, the binding energy between X and Y in the Fe matrix is[22];

$$E_{\_}(X,Y)(Fe) = E\big((N-1)Fe + 1X\big) + E\big((N-1)Fe + 1Y\big) - E\big((N-2)Fe + 1X + 1Y\big) - E(N.Fe), \quad X, Y = vacancy(\square), O, Al, Y, Zr \quad \text{----(B)}$$

If X and Y attract, binding energy will be negative and if they repel, binding energy will be positive.

Differential charge density plots of defect species in the Fe matrix are plotted. Differential charge density for a system AB is:

$$\Delta\rho = \rho_{AB} - \rho_A - \rho_B \qquad \text{---(C)}$$

where, $\rho_{AB}$ is the electronic charge density of the combined system, $\rho_A$ and $\rho_B$ being that of individual systems. In the contour plots, the minimum value of charge density is -0.006 e/Bohr$^3$ and the maximum value is +0.006 e/Bohr$^3$ with an interval 0.001 e/Bohr$^3$. The contour lines are plotted in a linear way such that the numerical value of the N$^{th}$ contour line is :

$$F(N) = Min + N \times Interval \qquad \text{---(D)}$$

The dashed line indicates negative contours, and solid lines indicate positive contours. Higher the density of lines, the change is drastic. Positive contours increase from periphery to center starting from zero to maximum by the interval mentioned. Similar way, negative contours decrease from zero to a minimum on moving towards the center.

### III. Results and Discussions

#### A. *Formation energies of solute atoms*

Formation energies are calculated for solute atoms in tetrahedral, octahedral and substitutional sites of bcc Fe supercell. It is observed that oxygen prefers the octahedral interstitial site in Fe matrix while Y, Al and Zr prefer substitutional position. Calculated formation energies for Y, Zr, Al, O and vacancy in defect-free Fe matrix and the existing data on the same are tabulated in Table.1. The formation energy of Al in the octahedral site, tetrahedral site and substitutional sites are +5.10 eV, +4.87 eV and -0.78 eV, respectively. So, Al prefers substitution position in the Iron matrix. Formation energies also give insight into the solubilities of elements in the matrix. In the bcc Fe matrix, up to 45 at% of Al can be dissolved depending on the temperature[23]. The solubility of Zr is limited to less than 0.16 at %[24, 25], and for Y, it is less than 1 at%[25]. The reported solubilities of oxygen in Fe has considerable discrepancies because of the strong influence of impurities like Si and Al on the solubility[26]. The purer the iron sample used, the lower the solubility of Oxygen. The tentative solubility as per [27] is 2.4× 10$^{-3}$ at%. So, more positive formation energy corresponds to less solubility and vice versa.

**Table.1.** Formation energies of Zr, Y, Al, and O in the bcc Fe matrix.

| Element | Formation Energy(eV) | Literature data |
| --- | --- | --- |

| Zr | +0.32 | +0.42 [28] |
| Y  | +1.90 | +1.86[13]   +1.72[29]   +2.02[30]   +2.12[31] |
| Al | -0.79 | −0.75[31] |
| O  | +1.30 | +1.35[13], +1.45[32], +1.41[30] |

The difference in charge density when Al, Zr, and Oxygen are introduced into the Fe matrix is plotted (Fig.1.(a, b, c, d)). The 2D slice shown is [1T0] plane of 128 atom bcc Fe supercell. Four first nearest neighbors and two second nearest neighbors of impurity atom are seen in the slice. In figure 1 (a), (b), and (c), there is a charge accumulation in the binding region as well as in the antibonding region. The accumulation of charge in an antibonding region is less compared to that in the bonding region. The charge is depleted from the immediate neighborhood of nuclei in a direction perpendicular to the bond direction forming a doughnut-shaped structure encircling the bond axis, which is the characteristic of covalent bonding [33]. The charge increase in the bonding region, responsible for the binding of nuclei, is shared by both the nuclei, and hence the interaction is covalent in nature.

Fe–O interaction, when Oxygen is in an octahedral void of Fe lattice, is ionic in nature. The magnetic nature of the Fe–O interaction is responsible for this. The magnetism of Fe lattice plays a role in confining the charge in the interstitial region[6]. If we assume that the bond is ionic, Fe and O exist as $Fe^+$ and $O^-$ ions, and then the condition for a bond to be stable is that the force on both nuclei should be zero. This is observed in charge density difference plot (Fig.1.d), and hence Fe–O interaction is ionic in nature.

The nearest neighbor distance table gives an idea of how much the neighboring atoms moved from their equilibrium position upon the introduction of the impurity atom. There are eight first-nearest neighbors, six second-nearest neighbors, and 12 third-nearest neighbors for an atom in bcc lattice. Being in an interstitial site, Oxygen creates maximum distortion.

**Table 2.** Nearest neighbor distances in Fe matrix with and without impurity atoms. (nn denotes the nearest neighbor).

|  | Pure Fe (Å) | With Zr atom(Å) | With Al atom(Å) | With O atom(Å) | With Y atom(Å) |
|---|---|---|---|---|---|
| 1 nn | 2.45 | 2.55 | 2.48 | 2.69 | 2.59 |
| 2 nn | 2.81 | 2.86 | 2.79 | 2.90 | 2.88 |
| 3 nn | 3.97 | 4.02 | 4.00 | 4.08 | 4.03 |

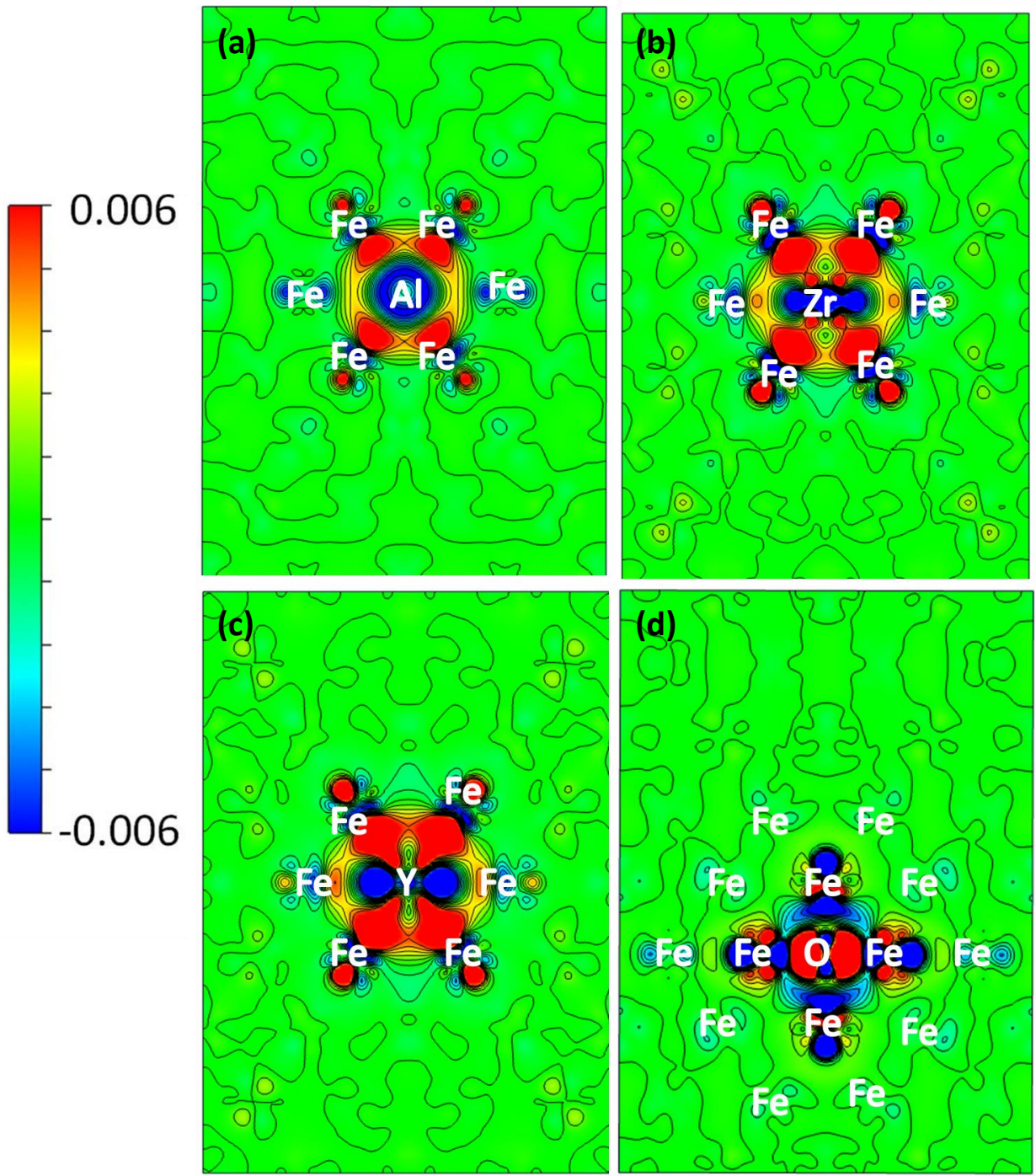

**Fig. 1.** The differential charge density plot of (a) Al, (b) Zr, (c) Y and (d) Oxygen in Fe matrix in [1T0] plane of 128 atom bcc Fe supercell. Blue colour is -0.006 e/Bohr³ and red is +0.006 e/Bohr³. Al, Y and Zr are in substitutional position while oxygen is in the octahedral interstitial site. Al, Y and Zr bind covalently with surrounding Fe atoms. Oxygen – Fe binding is ionic in nature, which is evident from the polarization of Fe-O bond.

### B. Solute – Solute Interaction:

Now, if we put two similar impurity atoms in the iron matrix in first nearest neighbor positions, the interactions are repulsive irrespective of the species considered. So in a nanocluster, it is unlikely that two metallic atoms will come close together (Table.3.). At the second nearest neighbor position, interaction for Zr with Zr is slightly attractive. For Y–Al it remains repulsive even for the second neighbor position.

**Table 3.** Binding energies of two metallic atoms in Fe matrix at first and second nearest neighbor (nn) positions.

| Defect Species | Formation Energy(eV) | | Literature data |
|---|---|---|---|
| | 1nn | 2nn | |
| Zr -Zr | 0.12 | -0.11 | |
| Y-Y | 0.13 | -0.06 | 0.19, -0.01[32]; 0.20, 0[30] |
| Al-Al | 0.12 | -0.03 | |
| Zr-Y | 0.16 | -0.04 | 0.12[13] |
| Zr-Al | 0.15 | -0.02 | |
| Y-Al | 0.24 | 0.11 | |

### C. Solute – Vacancy Interaction:

Solute vacancy interaction should be studied since the supersaturation of vacancies during mechanical alloying and their binding with solute atoms play a major role in nanocluster nucleation[6]. Also, vacancy-solute binding is crucial in determining the diffusion kinetics of impurity in a matrix[34]. The solute, which is least bound with Fe matrix, one with the largest positive formation energy, is expected to bind more with vacancy. The expected trend is followed for Zr, Al and Y in Fe matrix, Y–□ binding being the strongest (-1.20 eV) followed by Zr–□ binding (-0.78) and Al–□ binding energy (-0.30) being the least of these three (Table 4.). However, oxygen binds more with vacancy than all the above owing to its magnetic nature. The magnetism of host lattice causes confinement of O charge density in the interstitial region. The presence of vacancy reduces the charge confinement by creating a new volume for charge delocalization[6].

Yttrium and Zirconium interact repulsively with Fe atoms and their size is bigger than Fe. Such a large impurity introduces strain in the matrix. The vacancies near to such atoms will provide an option for them to move further away from the Fe atom relieving the strain. But, a vacancy next to a well-bound Al-Fe pair is unfavorable since Al atom is small and interaction with vacancy requires breaking a strong Fe-Al bond.

**Table 4.** The binding energy of Y, Al and Zr with a vacancy.

| Defect species | Binding Energy(eV) | | Literature data |
| --- | --- | --- | --- |
| | 1nn | 2nn | |
| Y -□ | -1.20 | -0.05 | -1.45, -0.26[32] |
| Al-□ | -0.30 | +0.05 | |
| Zr -□ | -0.78 | -0.06 | |
| O-□ | -1.60 | -0.62 | -1.45,-0.60[6] |

Charge density plots for the interaction of Zr and Al with vacancy are given in Fig.2.(a) and (b). In the presence of vacancy in the first nearest neighbor position to an Al atom in Fe matrix, the bonding with three first nearest neighbors remain the same (Fig.2.(b)). But in the case of Zr ((Fig.2(a)), as a vacancy is placed in the first nearest neighbor position, the bond diagonally opposite to the vacancy is broken. The charge is redistributed in such a way that the bond with the second nearest neighbor is strengthened.

The calculated nearest neighbor distances indicate that Zr has moved 0.3 Å towards vacancy, which makes the distance to the first nearest neighbor diagonally opposite to Zr atom as 2.85 C (which was previously 2.55 Å as per table 2). The distance to second nearest neighbors, which was 2.8 Å has reduced to 2.6 Å, and hence the bonds get rearranged accordingly. However, the displacement of the Al atom towards vacancy is only 0.1 Å.

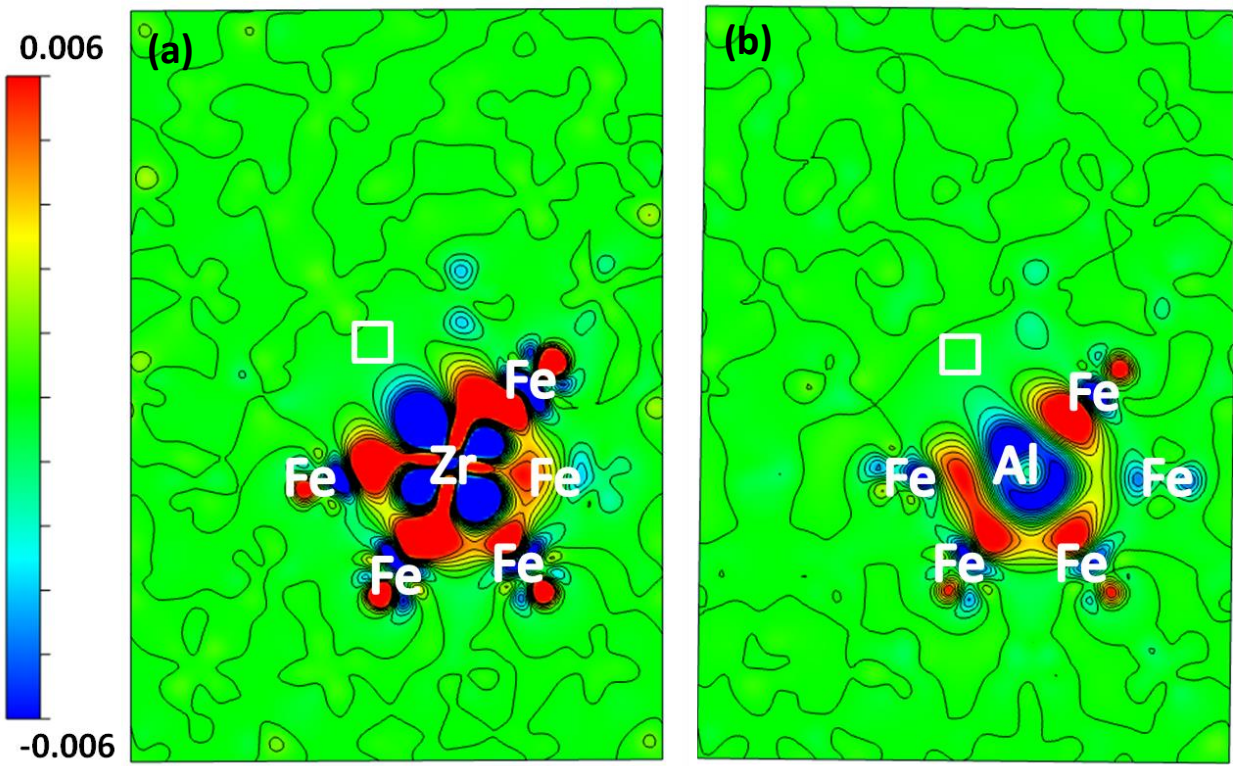

**Fig. 2.** [1T0] slice of charge density difference plot of the interaction of Zr and Al with a vacancy. (Vacancy is marked by a white square.)

### D. *Solute – Oxygen Interaction:*

Oxygen is the next important element of nanocluster. Calculated binding energies of Y, Al, Zr, and Vacancy with oxygen is given in Table.5. There is an appreciable binding between Al and Oxygen when they are first nearest neighbors. For Zr, the attraction is noticeable when they are at the second nearest neighbor position. In the free state, the bond energy of Zr–O, Al–O and Y–O is in the decreasing order ($E_{Zr-O} > E_{Y-O} > E_{Al-O}$)[35]. But the calculations show that inside Fe matrix, Y–O bond is stronger than Zr–O bond ($E_{Y-O} > E_{Zr-O} > E_{Al-O}$). In the defect-free Fe lattice, formation energies are of order $E_{Fe-Y} > E_{Fe-Zr} > E_{Fe-Al}$, which explains the anomaly.

**Table.5.** Solute – Oxygen binding energy in the bcc Fe matrix. □-O has the highest binding, followed by Zr-O and then Al-O.

| Element | Binding Energy(eV) | Literature data |
|---------|-------------------|-----------------|

|      | 1nn   | 2nn   |                      |
| ---- | ----- | ----- | -------------------- |
| Y -O | +0.28 | -0.91 | +0.35, -1.01[13]     |
|      |       |       | +0.28, -0.85[30]     |
| Al-O | -0.23 | +0.08 |                      |
| Zr -O| +0.01 | -0.80 |                      |
| □-O  | -1.60 | -0.62 | -1.45,-0.60[6]       |

Fig.3. shows [1T0] plane of Charge density difference plot of Zr–O interaction in the Fe matrix. It shows the ionic nature of interactions. The charge is polarized towards oxygen. The polarization perturbs Zr atom's bonding with neighboring Fe atoms. In order to balance the force on the Zr nucleus due to polarization of electrons towards O atom, Fe atoms up to fourth nearest neighbor are affected.

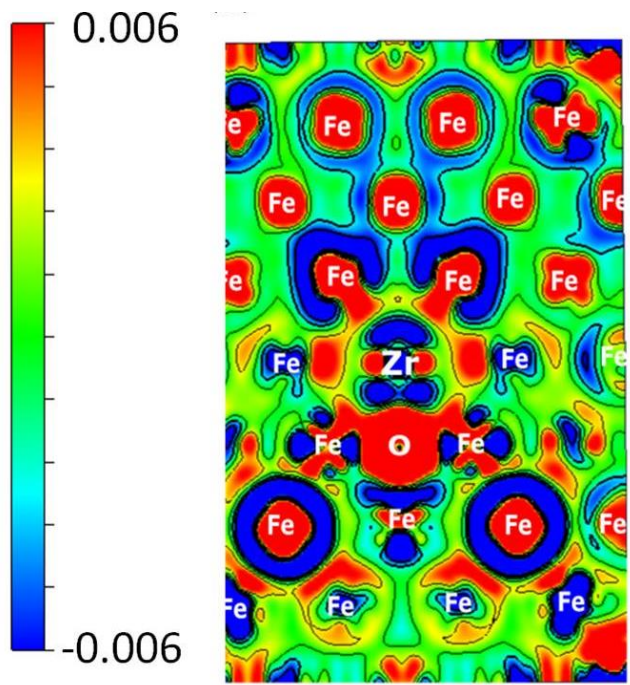

**Fig.3.** [1T0] plane of differential charge density plot of Zr with Oxygen in 128 atom Fe matrix.

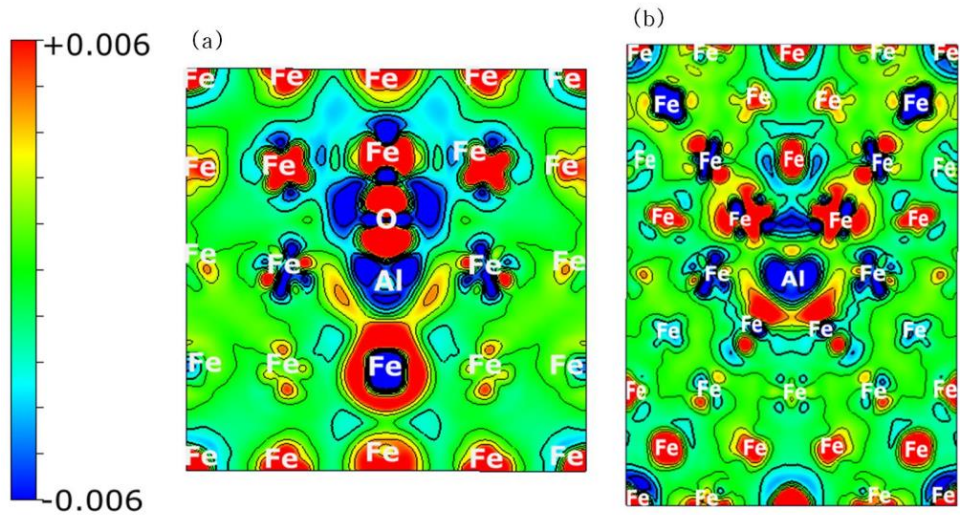

**Fig.4**. Differential charge density plot of Al with Oxygen in 128 tom Fe matrix. (a) [02$\bar{1}$] plane (b) [1$\bar{1}$0]plane

Charge density difference in [1$\bar{1}$0] plane of 128 atom supercell containing Al and oxygen in first nearest neighbor positions, given in Fig.4(b)., shows the effect of oxygen on Fe-Al bonding. The oxygen atom is in between Al and Fe above the plane. Out of four Fe-Al covalent bonds (see Fig.1(a)) otherwise present, two remains unaltered while two are broken. The [02$\bar{1}$] plane, where oxygen is in the plane, shows Fe-O and Al-O interactions, which are ionic in nature. The effect of polarization lasts till the fourth nearest-neighbor position, but the charge delocalization in atoms other than the next nearest-neighbors are less prominent compared to Zr–O interaction.

### E. *Solute – Vacancy - Oxygen Interaction:*

Further, Oxygen and Vacancy are placed next to the solute atom in the Fe matrix. The possible interactions are solute–oxygen, solute–vacancy, and oxygen–vacancy interactions. Out of four combinations possible, which are listed in Table.6., one with maximum binding energy is the most stable configuration. The maximum binding energy for Zr-O-□ and for Al-O-□ clusters are -2.55 eV and -2.01 eV, respectively. In the charge density plot of Zr–O–□ in Fe matrix (Fig.5), the charge delocalization is less compared to Zr–O plot (Fig.3.) It is evident that the presence of vacancy helps to confine the charge delocalization within second nearest neighbors. When Al-□-O is placed in Fe matrix (Fig.6), the net binding energy of cluster is less than Zr–O–□ binding energy because Al–□ binding energy is lesser than Zr–□ binding energy.

**Table.6.** Binding energies of different Solute – Oxygen – Vacancy combinations in bcc Fe. In the configurations shown, the Green circle is solute (Zr or Al) atom, red is oxygen atom and white is Fe atom. The vacancy is indicated by a cube.

| Defect Configuration | Binding Energy(eV) | |
|---|---|---|
| | Zr | Al |
| 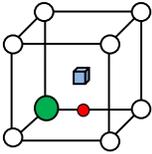 | -0.6 | -0.94 |
| 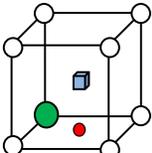 | -2.55 | -1.68 |
| 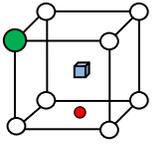 | -2.30 | -2.01 |
| 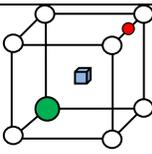 | -1.35 | -0.90 |
| 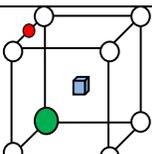 | -1.4 | -0.93 |

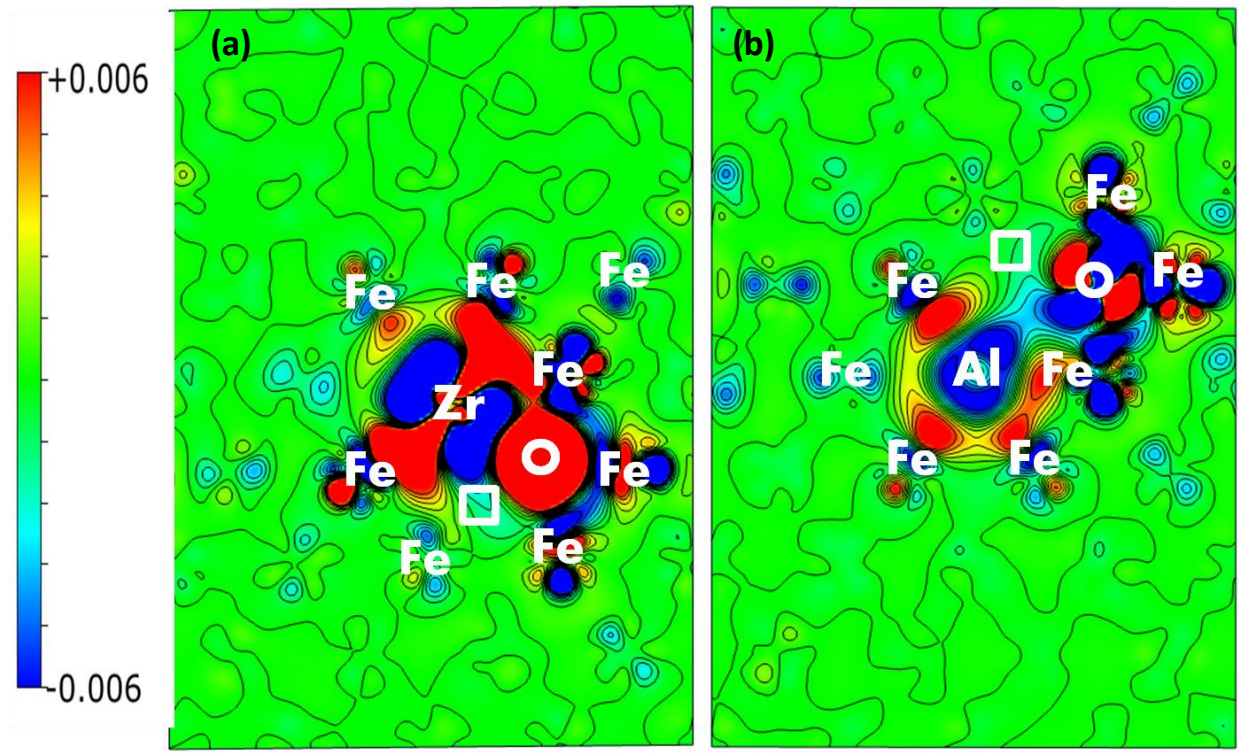

**Fig.5**. The [1T0] plane of charge density difference plot of the interaction of (a) Zr–O–Vacancy cluster and (b) Al–O–Vacancy cluster in 128 atom supercell of Iron. The vacancy is denoted by a white square.

### F. Yttrium - Solute – Vacancy - Oxygen Interaction:

Further, all the known constituents of a nanocluster, Yttrium. Oxygen, vacancy, and solute atom are introduced in the Fe matrix. Combinations of these with negative binding energies are listed in Table 7. Maximum binding energy is -3.25 eV for Y–O–□–Al cluster and -4.13 eV for Y–O–□–Zr cluster. The [0$\bar{1}$1] plane of charge density plots for maximum binding energy configurations of both clusters are shown in Fig.6. It contains three O-Fe, one O-solute, one Y-O ionic bonds, and three solute –Fe and two Y–Fe bonds of covalent nature. Charge density difference plots show that charge distortion in the matrix is minimized due to the presence of vacancy.

**Table.7.** Binding energies of different yttrium–solute– Oxygen–Vacancy combinations in bcc Fe. green is solute, the orange circle is yttrium, the red is Oxygen and cube is the vacancy.

| Defect Configuration | Binding Energy(eV) | |
| --- | --- | --- |
| | Zr | Al |
| 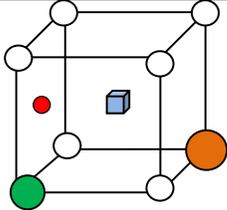 | -3.89 | -3.03 |
| 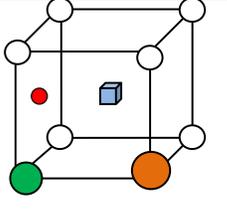 | -3.25 | -2.81 |
| 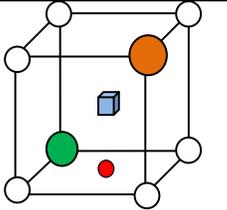 | -4.13 | -3.15 |
| 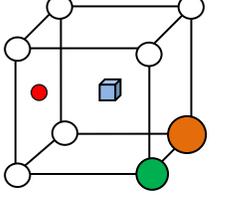 | -2.86 | -2.94 |
| 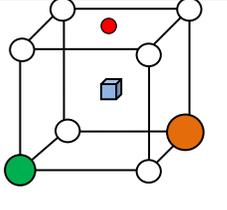 | -3.74 | -3.25 |
| 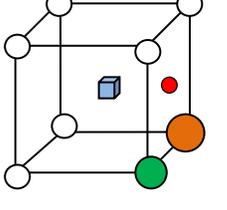 | -3.54 | -2.66 |
| 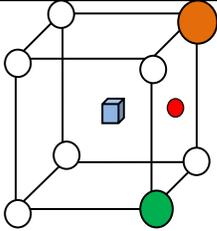 | -3.55 | -2.82 |
| 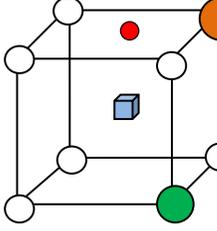 | -3.57 | -2.97 |
| 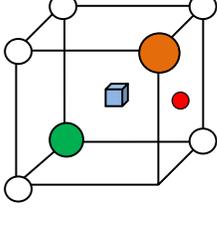 | -3.66 | -3.01 |
| 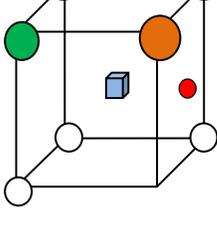 | -3.06 | -2.73 |

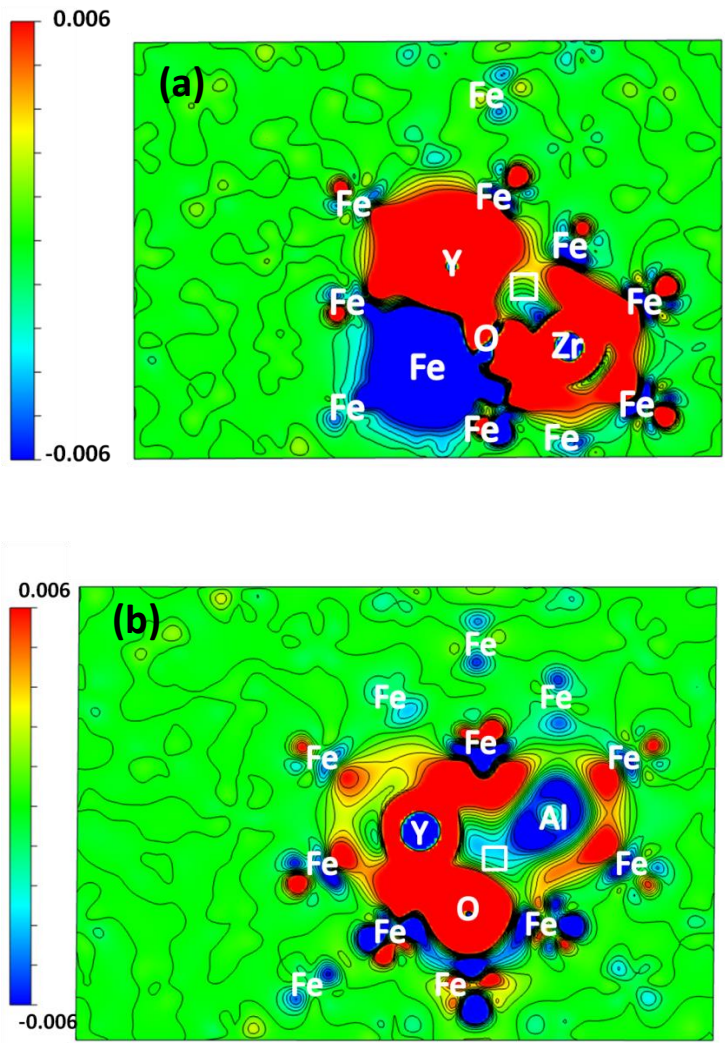

**Fig.6.** (a) Charge Density Difference Plots of maximum binding energy configuration of Y- Zr –O –Vacancy cluster in [0$\bar{1}$1] plane of 128 atom supercell.(b) Charge Density Difference Plots of maximum binding energy configuration of Y- Al –O –Vacancy cluster in [0 $\bar{1}$1] plane of 128 atom supercell.

## IV.    Conclusions

In the present work, The formation energies of relevant elements, their binding with Yttrium oxygen, vacancy, O–Vacancy and Y–O–Vacancy in bcc Fe matrix are found using ab initio calculations. The binding energies in the various stages of cluster formation is summarized in the graph (Fig.7.) It is clear that the Zr containing clusters have higher binding energy compared to Al containing clusters for all the selected configurations, which means the presence of Zr and Al together in a ferrite matrix with Yttria and Oxygen will support the formation of Y-Zr-O clusters than Y-Al-O clusters. Thus it justifies the experimental finding of Y–Zr–O nano particles in large number compared to Y–Al–O nano particles in Oxide dispersion strengthened steels containing both Zr and Al[10].

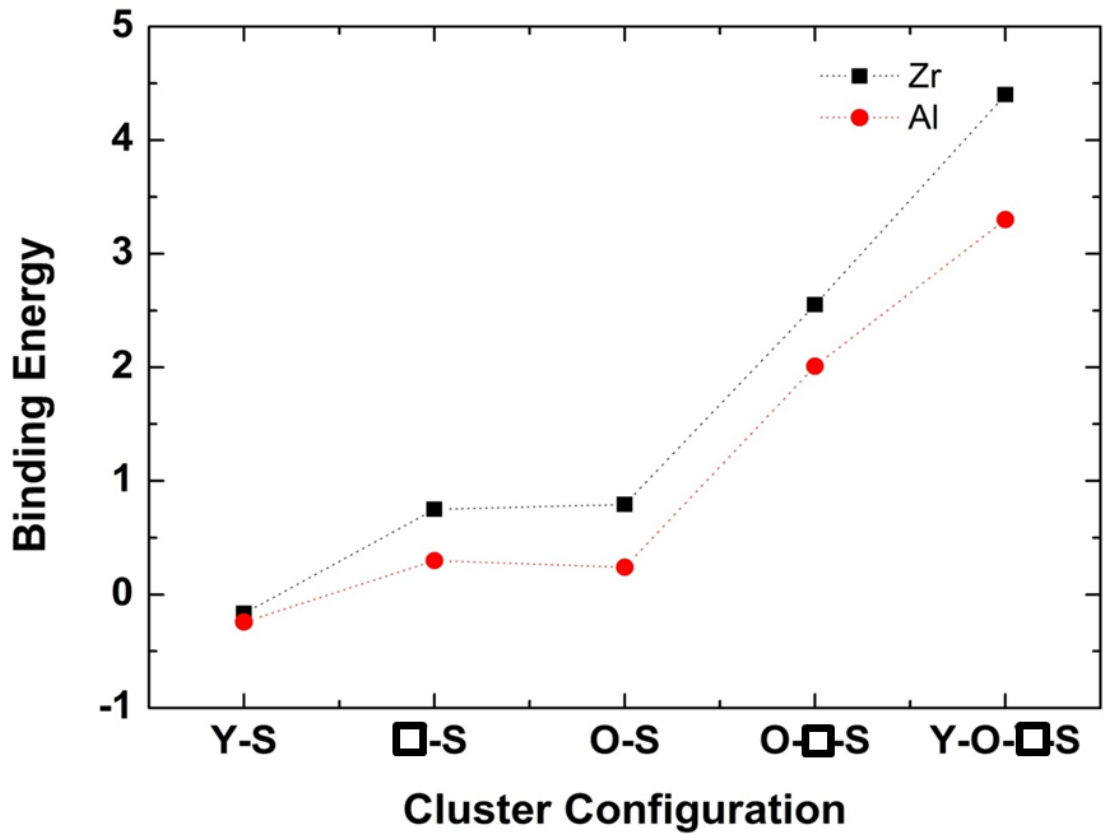

**Fig.7.** Absolute values of binding energies of Zr and Al containing clusters in bcc Fe (here S is solute: Zr/Al). The binding energy of Zr containing clusters are always higher than that of Al containing clusters.